\documentclass[prl,showpacs,superscriptaddress,preprintnumbers,amssymb,twocolumn,amsfonts]{revtex4-1}
\usepackage{graphicx}
\usepackage{amssymb}
\usepackage{amsmath}
\usepackage{subfigure}
\usepackage{bm}

\begin{document}

\def\beq{\begin{equation}}
\def\eeq{\end{equation}}
\def\beqn{\begin{eqnarray}}
\def\eeqn{\end{eqnarray}}
\def\etal{\emph{et al.}}
\def\ket#1{\vert #1 \rangle}
\def\bra#1{\langle #1 \vert}
\def\ev#1{\langle #1 \rangle}
\def\ip#1#2{\langle #1 \vert #2 \rangle}
\def\me#1#2#3{\langle #1 \vert #2 \vert #3 \rangle}
\renewcommand{\bf}{\mathbf}

\title{Edge excitations of bosonic fractional quantum Hall phases in optical lattices}

\author{Jonas~A.~Kj\"{a}ll}
\affiliation{Department of Physics, University of California,
Berkeley, CA 94720}
\author{Joel~E.~Moore}
\affiliation{Department of Physics, University of California,
Berkeley, CA 94720} \affiliation{Materials Sciences Division,
Lawrence Berkeley National Laboratory, Berkeley, CA 94720}
\date{\today}

\begin{abstract}
The rapid development of artificial gauge fields in ultracold gases suggests that atomic realization of fractional quantum Hall physics will become experimentally practical in the near future.  While it is known that bosons on lattices can support quantum Hall states, the universal edge excitations that provide the most likely experimental probe of the topological order have not been obtained.  We find that the edge excitations of an interacting boson lattice model are surprisingly sensitive to interedge hybridization and edge-bulk mixing for some confining potentals.  With properly chosen potentials and fluxes, the edge spectrum is surprisingly clear even for small systems with strong lattice effects such as bandwidth.  Various fractional quantum Hall phases for bosons can be obtained, and the phases $\nu=1/2$ and $\nu=2/3$ have the edge spectra predicted by the chiral Luttinger liquid theory.

\end{abstract}
\maketitle

Fractional quantum Hall (FQH) phases~\cite{Tsui,DasSarmabook} contain a wide variety of interesting physics, including topologically degenerate ground states, fractional bulk excitations, and gapless chiral edge excitations. They arise at low temperatures when strong magnetic fields are applied to high-quality two-dimensional electron gases with low carrier concentration.  Ultracold gases of neutral atoms are being used to investigate several properties of materials which can be hard to control precisely in the solid state.  As these systems are charge-neutral, an ordinary magnetic field cannot be used to create the Lorentz force.  A synthetic magnetic field can be created by rotation, but technical issues appear to limit this approach to lower field strengths than are necessary for FQH~\cite{Spielman}, with the exception of a recent experiment  with few trapped particles~\cite{Gemelke}.  Much theoretical work has been done for these systems, for a review see~\cite{Cooperrev}, including edge spectrum calculations~\cite{Cazalilla}.    

Recently, several theoretical~\cite{Sorensen,Dalibard,Cooper} and experimental~\cite{Spielman,Bloch} proposals have been made for stronger synthetic magnetic fields for ultracold neutral atoms. All of them can be used with optical lattices which enhance interaction effects and give a larger energy gap above the FQH ground state. Theoretical work on lattice systems with an effective magnetic field goes back at least to Hofstader's work~\cite{Hofstadter} on non-interacting particles; the FQH phases are strongly interacting, and S{\o}rensen et~al.~\cite{Sorensen} showed that in the low flux limit and strong interactions the system could be well described by Laughlin's wavefunction~\cite{Laughlin}. In subsequent work Hafezi et~al.~\cite{Hafezi} concluded that this could be extended to larger fluxes per unit cell by investigating the topological structure of the ground state.

The goal of this work is to understand practical experimental conditions for observation of edge states in bosonic lattice FQH systems and compare numerical results for edge excitations in hierarchy states to the prediction of chiral Luttinger liquid theory~\cite{Wenbook}.  Convincing observation of bosonic FQH states will depend on an experimentally viable probe of the topological order; while many multi-particle quantities have been used to diagnose the topological state in past theoretical work, such as ground state degeneracy, bulk energy gap, wavefunction overlap, band flatness, band Chern number and entanglement spectra, these are not yet experimentally accessible.  Our focus will be on edge excitations, whose ``universal'' aspects contain information about the topological order of the system, although a good understanding of the ``non-universal'' effects of the lattice and trap is crucial for these excitations to provide a clear signal that a FQH phase has been obtained.  

We investigate bosonic FQH phases in the simplest lattice system, hard-core bosons on a square lattice in a uniform magnetic field with equivalent Landau level filling $\nu=N/N_{\phi}$. $N$ is the number of bosons and $N_{\phi}$ is the number of fluxes in the system, measured in units of the magnetic flux quantum $\Phi_0=hc/q$, for particles of charge $q$. The modified Bose-Hubbard Hamiltonian is, 
\begin{equation}\label{eq:Ham}
H=-J\displaystyle\sum_{\vec{r}}\hat{a}^{\dagger}_{\vec{r}+\hat{x}}\hat{a}_{\vec{r}}e^{-i\alpha_xy}+\hat{a}^{\dagger}_{\vec{r}+\hat{y}}\hat{a}_{\vec{r}}e^{i\alpha_yx}+h.c.,
\end{equation}
where $J$ is the hopping amplitude, $\hat{a}^{\dagger}_{\vec{r}}$ creates a boson on site $\vec{r}=(x,y)$.  We use two different gauges for the phases $\vec{\alpha}=$($\alpha_x,\alpha_y$), Landau gauge $\vec{\alpha}=(\alpha,0)$ on cylinders to keep explicit translational symmetry around their circumferences, and symmetric gauge, $\vec{\alpha}=(\alpha/2,\alpha/2)$ on squares to keep explicit ${\mathbb Z}_4$ rotational symmetry. The flux through a plaquette $n_{\phi}=\alpha/2\pi$ is defined modulo 1 and can be expressed as an artificial magnetic field $\vec{B}^*=n_{\phi}\Phi_0/a^2\hat{n}$, where $a$ is the lattice spacing and $\hat{n}$ the vector normal to the lattice plane.  At low flux $n_{\phi}\ll 1$, a continuum description can be used and the system is effectively in the flat band limit~\cite{Hormozi}.  We focus on larger fluxes where the lattice is important.  The magnetic length $l_B$ is of the same order as the lattice spacing for the fluxes we are interested in, $l_B=\sqrt{\hbar c/(qB^*)}=a/\sqrt{2\pi n_{\phi}}\sim a$.

\begin{figure}[tbp]
  \begin{center}
    \subfigure{\includegraphics[width=38mm]{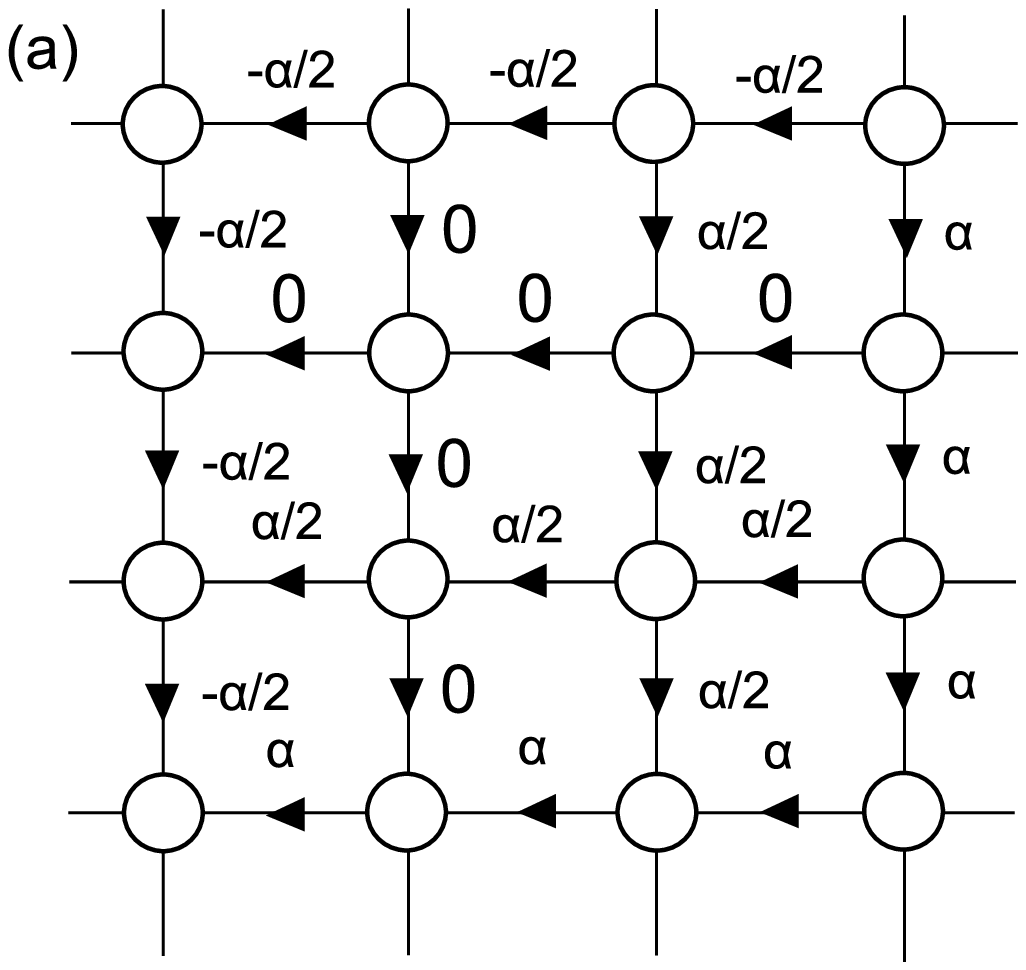}}
        \subfigure{\includegraphics[width=42mm]{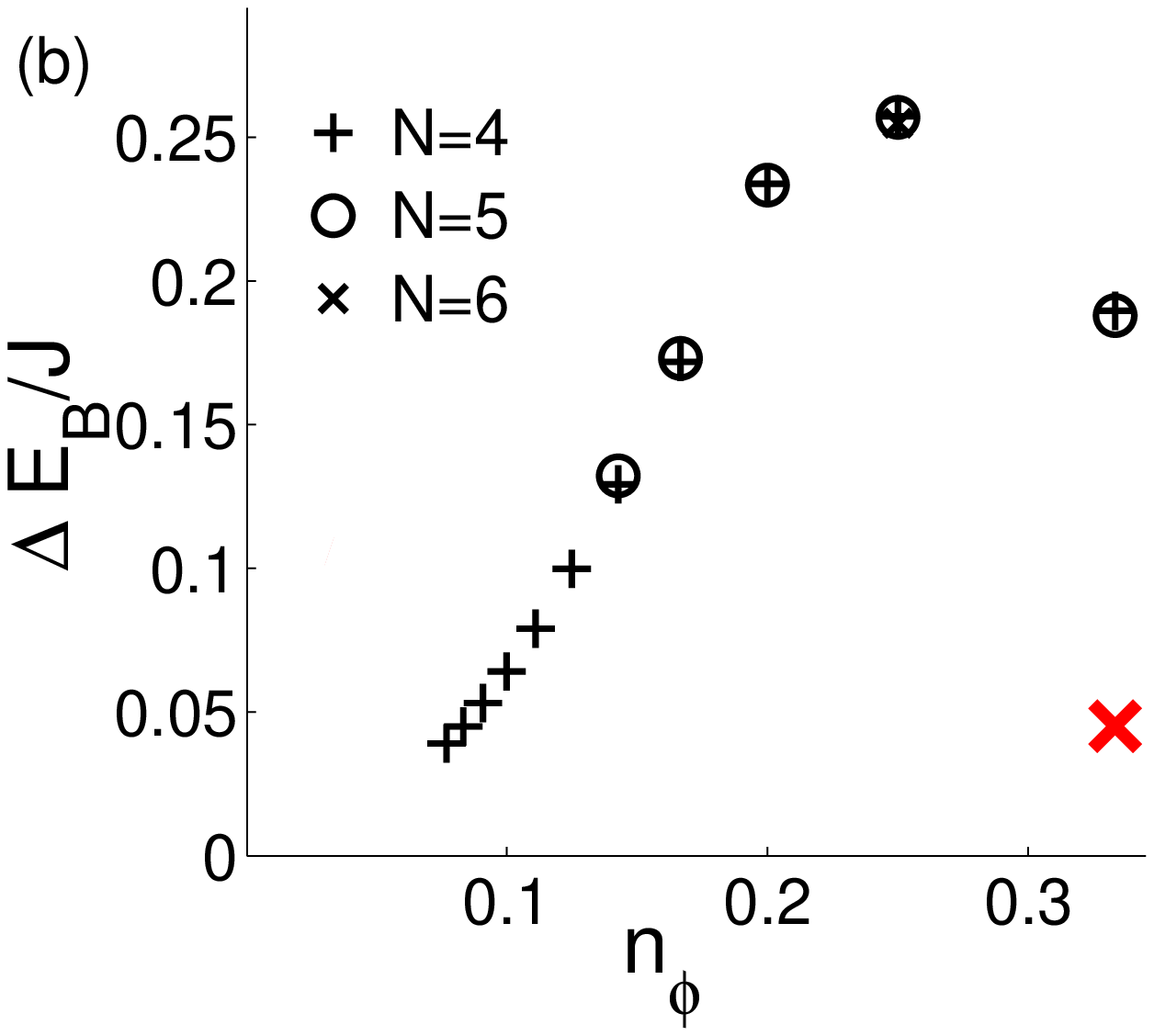}}
    \caption{(Color online) (a) Part of the square lattice in the symmetric gauge with the phases gained when hopping in the direction of the arrows. When hopping in the opposite direction, the phases are the complex conjugate of those shown. (b) The bulk gap in the $\nu=1/2$ phase on a cylinder as a function of flux per plaquette, for $N=4-6$ particles.  The large red cross is from a clearly different spectrum, indicating that it might not be in a FQH phase.}
    \label{Intro}
  \end{center}
\end{figure}

The spectrum of the Hamiltonian~(\ref{eq:Ham}) is computed with exact diagonalization. A section of the system is shown schematically in Fig.~\ref{Intro}(a). The edge excitations in an infinite system are gapless, but become gapped in a finite system.  To clearly see them, it is desirable to have a large bulk gap $\Delta E^B_{\nu}/J\equiv (E^B_{\nu}-E^{GS}_{\nu})/J$, where $E^B_{\nu}$ is the energy of the lowest bulk excitation and $E^{GS}_{\nu}$ the ground state energy in the $\nu$ phase.  The bulk gap on cylinders at $\nu=1/2$, in flat infinite wells where edge modes do not exist, are shown in Fig.~\ref{Intro}(b).  The ground state is non-degenerate and the gaps to all excited states are comparable to those reported for a torus~\cite{Sorensen}. 
To get edge excitations in the spectra, more sites outside the ground-state droplet need to be added where the edge waves can propagate. A trapping potential is then essential to confine the condensate. 

The optical trapping potential $V_{\text{trap}}(x,y)$, provides the equivalent of the electric field confinement in semiconductors. In this paper, we use harmonic traps $V_{trap}(x,y)=c_xx^2/a^2+c_yy^2/a^2$ to get one edge on squares and two edges on cylinders, with $c_x=0$ in the periodic direction of the cylinder.  The velocity of a non-interacting particle near the edge is $v=|\nabla V_{\text{trap}}(r_c)|/(n_{\phi}hc/a^2)$, where $r_c$ is the radius of the ground state droplet.  The edge excitations of FQH phases form chiral Luttinger liquids.  In the hydrodynamical approach~\cite{Wenbook}, the Hamiltonian of the edge waves in the Laughlin phases $\nu=1/m$, with $m=2,4,6,...$ for bosons, is
\begin{equation}\label{EdgeH}
H_{1/m}=2\pi\frac{v}{\nu}\displaystyle\sum_{k>0}\rho_{-k}\rho_k,
\end{equation}
with $[\rho_k,\rho_{k'}]=\frac{\nu}{2\pi}k\delta_{k+k'}$ where $\rho_k=L_e^{-1/2} \int d\theta e^{ik\theta /\hbar}\rho(\theta)$, $k=ph/L_e$ with $\rho(\theta)$ the one-dimensional density along the edge, $p\in{\mathbb N}$ and $L_e$ is the length of the edge. This is the $U(1)$ Kac-Moody algebra, describing a set of $k$ uncoupled harmonic oscillators with energy $\sum_pl_pv k$ and momentum $\sum_pl_p k$, where $l_p\in{\mathbb N}$ counts the number of excitations in each mode.  For a single edge, the degeneracies of the edge spectrum are $1,1,2,3,5,7,11,15,22,...$, see Tab.~\ref{tabell} for a labeling of the different states.  On a cylinder, there are two edges with momentum in different directions, right moving (R) and left moving (L). The degeneracies of this edge spectrum are $1,2,5,10,20,...$ or $(1),(1,1),(2,1,2),(3,2,2,3),(5,3,4,3,5),...$ if momentum is also resolved (Tab.~\ref{tabell}).  The wavefunctions in the microscopic theory of edge states for the Laughlin phases are
\begin{equation}\label{eq:EdgeL}
\Psi_{1/m}(z_i)=P(z_i)\displaystyle\prod_{i<j}(z_i-z_j)^{m}e^{-\displaystyle\sum_i|z_i|^2/4l_B^2},
\end{equation}
where $P(z_i)=\sum_p(\sum_i z_i^{p})^{l_p}$.  This is the form of all zero energy wavefunctions without a trap.  For a small number of particles, the edge excitations consisting of a few single particle modes $n_p=\sum_p l_p$ extends a distance $\Delta r\lesssim n_pl_B$ outside the ground state droplet. 

\begin{table}[tbp]
  \begin{center}
    \begin{tiny}
    \begin{tabular}{|c|c|c c c c c|}
	\hline & One edge & & & Two edges & & \\
        \hline E & p$=EL_e hv$& p$=0$ & p$=1$ & p$=2$ & p$=3$ & p$=4$ \\
	\hline 0 & GS & GS & & & & \\
	$hv/L_e$&1&  &1R & & & \\
	2$hv/L_e$& 2 & & & 2R & & \\
	&11& 1R1L & & 11R& & \\
	3$hv/L_e$& 3 & & & & 3R & \\
	& 21 & & 2R1L & & 21R & \\
	& 111 & & 11R1L & &111R& \\
	4$hv/L_e$& 4 & & & & &4R\\
	& 31 & & & 3R1L & &31R\\
	& 22 & 2R2L & & & &22R\\
	& 211 & 2R11L+11R2L & & 21R1L & &211R\\
	& 1111 & 11R11L & & 111R1L & &1111R\\ \hline
    \end{tabular}
    \end{tiny}
    \caption{Labeling of the edge spectra for one and two edges: $2\equiv\{l_2=1,l_i=0$ if $i\neq 2\}$, $111\equiv\{l_1=3,l_i=0$ if $i\neq 1\}$ and so on.  All edge levels with energy $E\le$4$hv/L_e$ and positive momentum $p\ge 0$ are shown.  For two edges, the levels with $p<0$ are equivalent to those with $p>0$, with L and R exchanged.}
    \label{tabell}
  \end{center}
\end{table}	

\begin{figure}[tbp]
  \begin{center}
    \subfigure{\includegraphics[width=41mm]{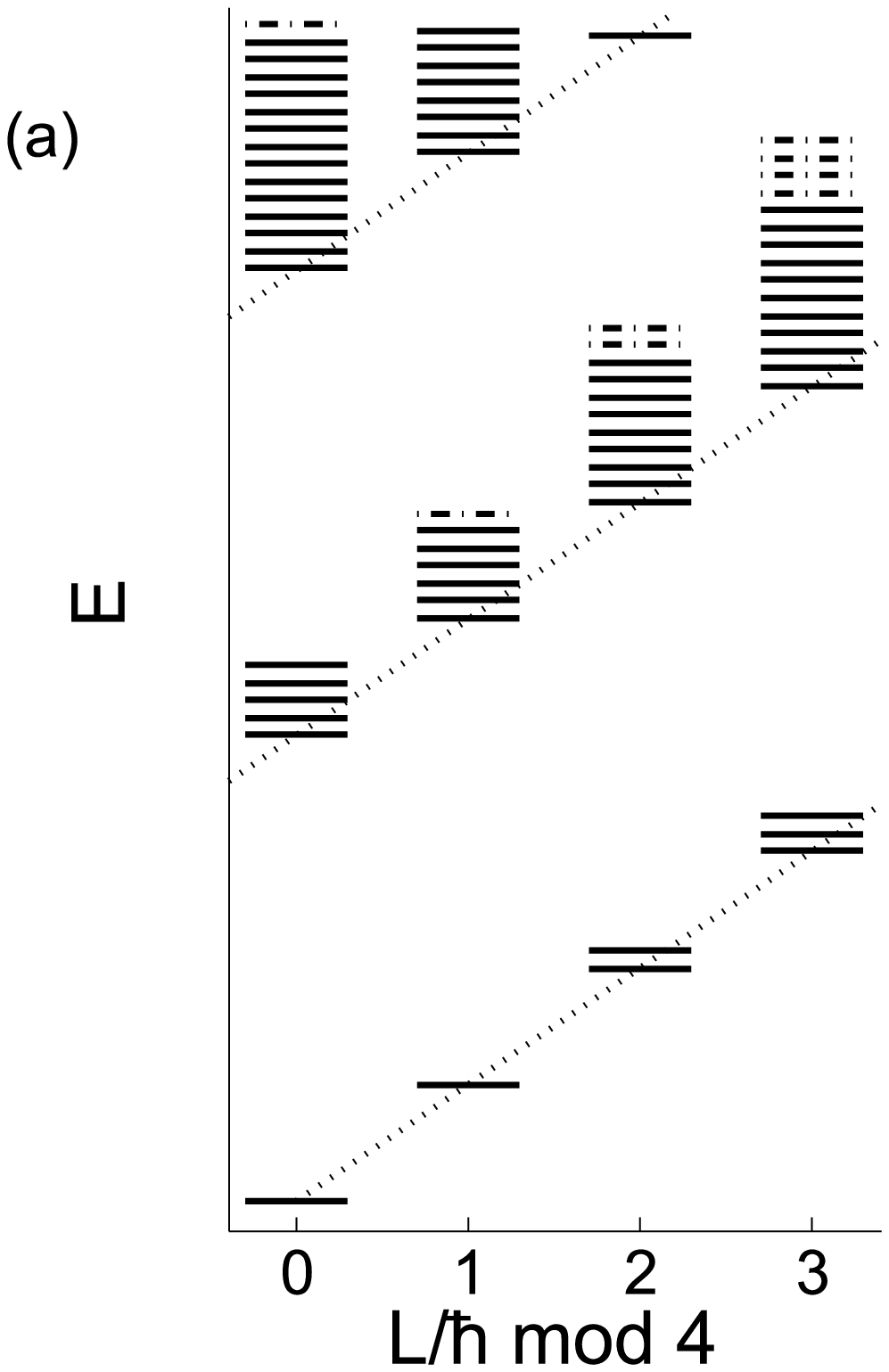}}
        \subfigure{\includegraphics[width=41mm]{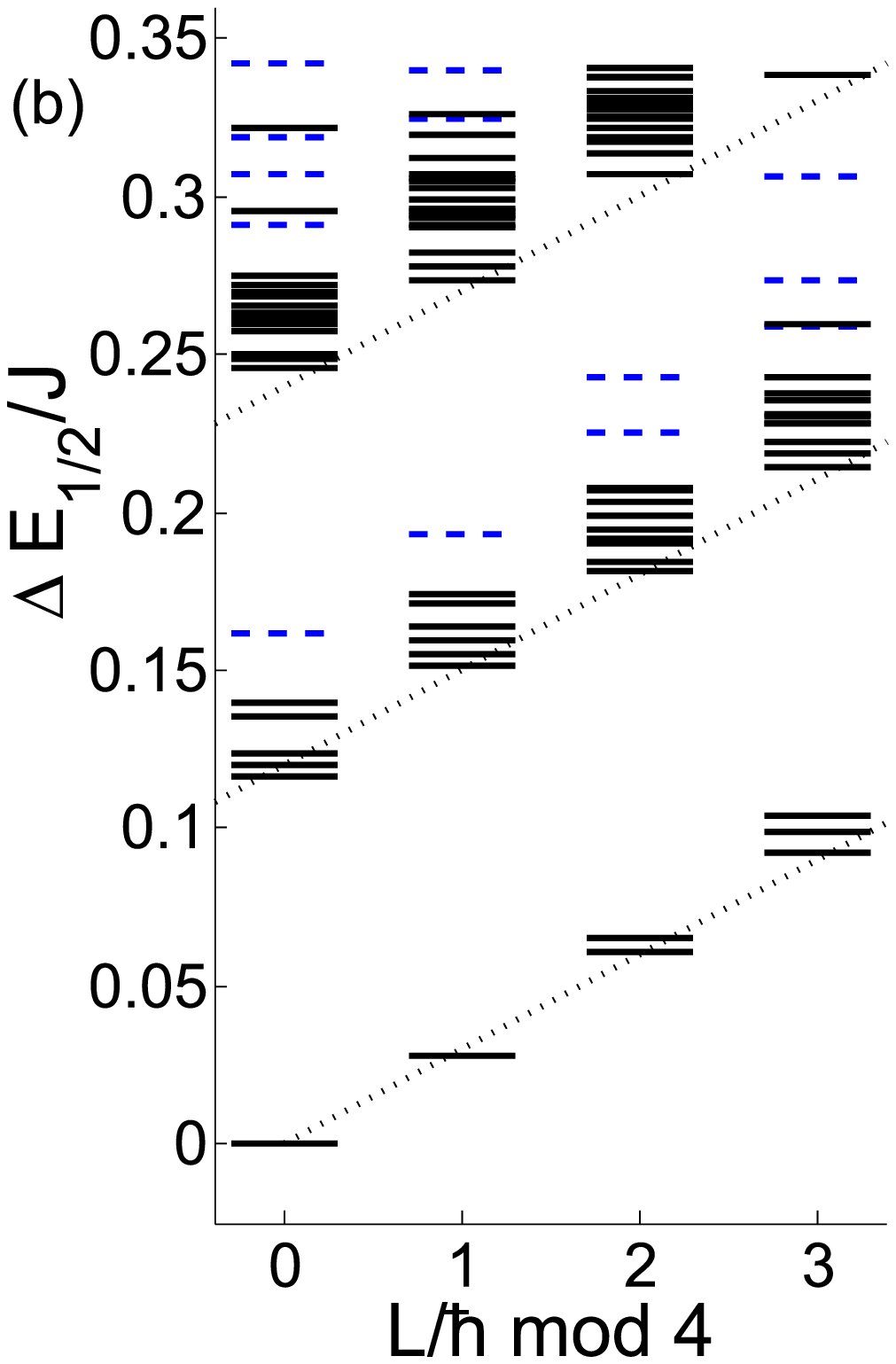}}\\
            \subfigure{\includegraphics[width=41mm]{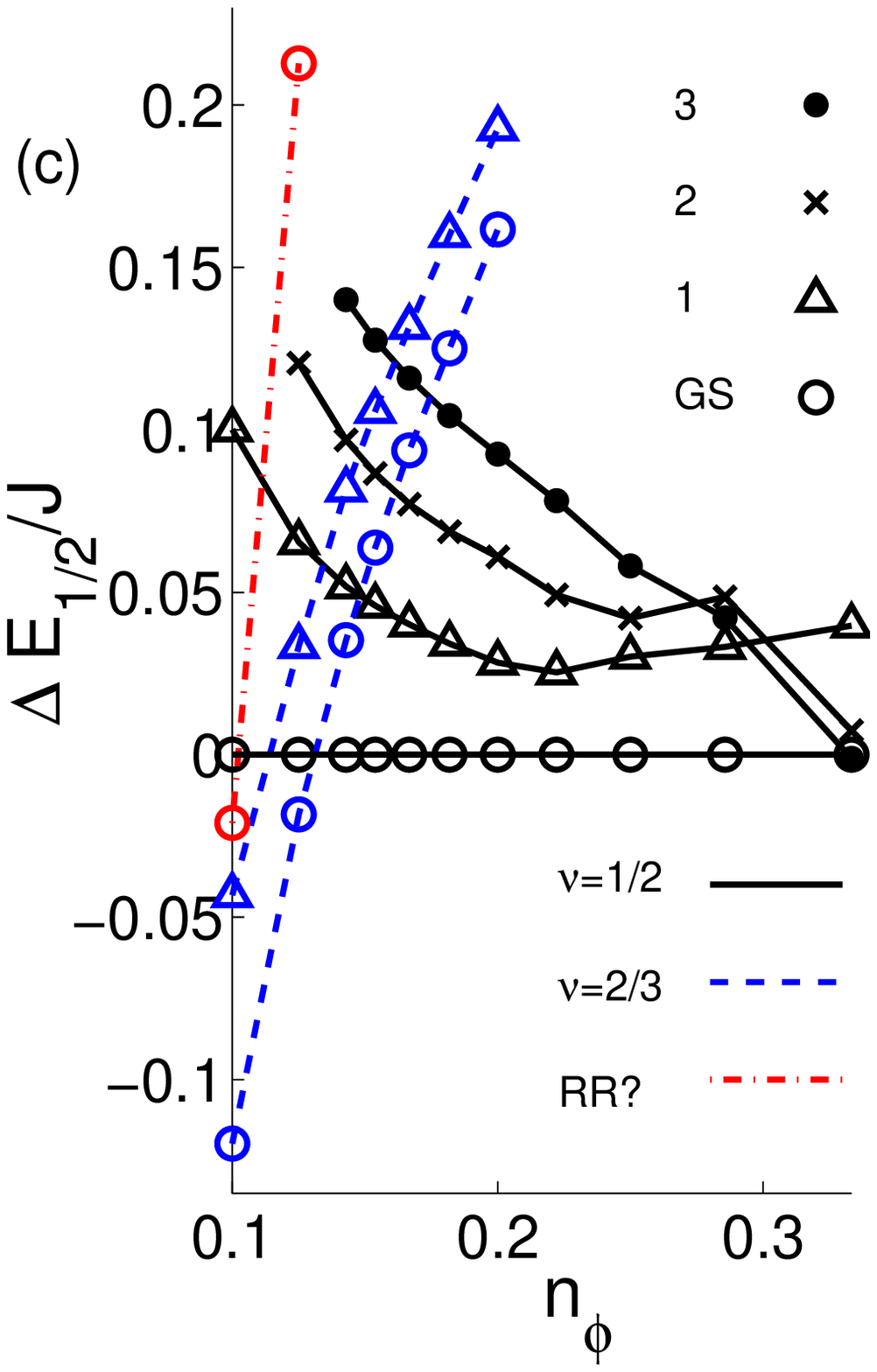}}
                \subfigure{\includegraphics[width=41mm]{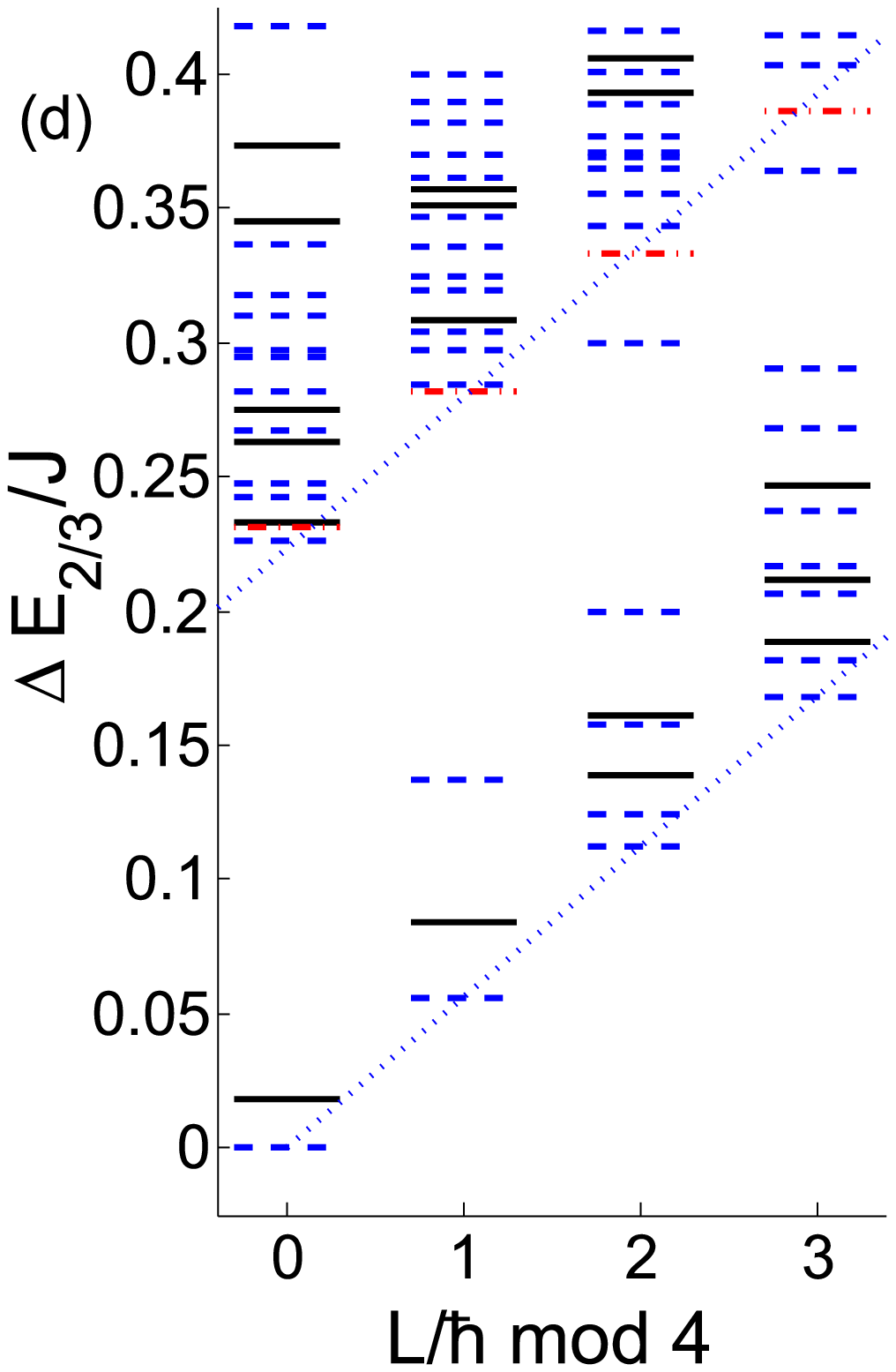}}
    \caption{(Color online) (a) Analytic edge spectrum at $\nu=1/2$ in a trap as a function of angular momenta.  Degenerate lines are drawn slightly apart for clarity.  The solid lines corresponds to excitations consisting of four single particle modes or less.  The dashed dotted lines are the additional excitations appearing for additional modes.  (b)-(d) Edge excitations for $N=4$ particles on a $11\times 11$ square lattice in a circular harmonic trap, $\nu=1/2$ black solid lines, $\nu=2/3$ blue dashed lines and the possible Read-Rezayi (RR?) phase, red dashed dotted lines. (b) Edge spectrum at $n_{\phi}=1/5$ as a function of angular momentum. (c) Energy gap to the $\nu=1/2$ ground state for: ground states ($\circ$), 1 states ($\bigtriangleup$), 2 states ($\times$) and 3 states ($\bullet$), labeled as in Tab.~\ref{tabell}, as a function of flux per plaquette. (d) Edge spectrum at $n_{\phi}=1/8$ as a function of angular momentum.}
    \label{Sq}
  \end{center}
\end{figure}

First, we consider the circular harmonic trap on a square lattice.  With an appropriate construction of the ${\mathbb Z}_4$ symmetric Hamiltonian, a Fourier transform will turn the Hamiltonian into block diagonal form with each block corresponding to a certain angular momentum $L/\hbar =0,1,2,3,...$ mod $4$.  These momenta $k_c=\Delta L/r_c$, with $\Delta L\equiv L-L^{GS}$, are the same as in the Kac-Moody algebra $k=ph/(2\pi r_c)$. The analytic edge spectrum for the $\nu=1/2$ phase is shown in Fig.~\ref{Sq}(a).  In a system with finite number of particles $N$, we only expect excitations consisting of $n_p\le N$ single modes.  The energy is proportional to the angular momentum for the trapped phase, indicated by the straight dotted line. With only $N=4$ particles on a $11\times 11$ site lattice in a trap $c_a/J=0.02$, we get remarkably good edge spectra.  At $n_{\phi}=1/5$, the filling fraction of the ground state is $\nu=1/2$. The degeneracies $1,1,2,3,5,6,9,11,15,...$ in the edge spectrum, agreeing with  $n_p\le N$ for $\nu=1/2$, is clearly visible, see Fig.~\ref{Sq}(b).  A straight line is a good fit to the lowest excitation at each momenta.  The small energy splitting is due to finite size effects, most noticable in the states that extends furthest in the trap.  Note that all of these state appear to be edge states; there are no bulk excitations for $\Delta E_{1/2}/J\equiv (E_{\nu}-E^{GS}_{1/2})\lesssim 0.35$, where $E_{\nu}$ is the energy of a state in the $\nu$ phase, much larger than the anticipated gap $\Delta E^B_{1/2}/J \approx 0.23$.  

Possibly the easiest way to experimentally measure the edge spectra is with stimulated two-photon Bragg scattering~\cite{Stenger}.  The probe light is scattered when the energy and momentum difference between the two beams is resonant with an excitation level in the FQH system.  
Inserting typical experimental values $J=\hbar/\tau_{tunnel}$, with a tunneling time $\tau_{tunnel}=0.2$ ms and $a=400$ nm gives an edge velocity $v=\Delta E_{\nu}/(\Delta L/r_c)\approx 0.2$ mm/s.  Resolving the closest energy excitations at the same angular momenta might initially be challenging.  However, the linear relation $v$ should be well within experimental reach with a long pulse duration $\delta t\sim 10$ ms in the Bragg spectroscopy during which the condensate has to remain stable.  
Modulation of the trapping potential is another commonly used technique to produce small energy excitations~\cite{Mewes}.  Detecting edge excitations could potentially
also be done by mapping out the in-situ density profile of
the condensate and an excited state, to observe the propagating
edge mode, or performing the analog to tunneling interferometry in condensed matter, either between the edges of
two condensates trapped next to each other [17] or by
outcoupling atoms from opposite points of the con-
densate. 

There are several states [Fig.~\ref{Sq}(b)] that do not belong to the $\nu=1/2$ edge spectrum. The lowest of those states has a $\nu=2/3$ filling factor.  The $\nu=2/3$ FQH phase has a 3-fold degeneracy on a torus and M\"{o}ller et~al.~\cite{Moller} showed it has a good average overlap with the composite-fermion wavefunction for $n_{\phi}\lesssim 0.3$ on a lattice.  Varying the magnetic flux slightly, states in the same phase change their energy in a similar manner, see Fig.~\ref{Sq}(c) for some examples.  
At small flux per plaquette, our system can become too small for some of the excited states of $\nu=1/2$ to exist.  Upon decreasing the flux further, other phases appear with larger ground state filling factors, consistent with some of the phases in the Read-Rezayi sequence for bosons $\nu=g/2$, where $g\in{\mathbb N}$~\cite{Read}.  Whether these phases actually are FQH phases are left for future studies.  Again, around $n_{\phi}\approx 0.3$, the edge spectrum break down and it is unclear what phases exist for $n_{\phi}\gtrsim 0.3$. 

The $\nu = 2/3$ hierarchical state~\cite{Haldane}, consists of two condensates with comoving edge modes at different radii.  
This spectrum has the same degeneracies as two edges on a cylinder $1,(1,1),(2,1,2),(3,2,2,3),...$, but states within each set () now have the same angular momentum, not energy. The edge spectra in Fig.~\ref{Sq}(d) is at $n_{\phi}=1/8$, where the ground state energy for $\nu=2/3$ is slightly lower than for $\nu=1/2$.  The $\nu=1/2$ edge excitations are spaced further apart, but the states with degeneracies $1,1,2,3,5$ can still be clearly distinguished. The two $U(1)$ branches of $\nu=2/3$ have very different velocitiess, but the structure $1,2,5$ can easily be seen under the dotted line corresponding to the lowest energy for the higher angular momentum states. 

Next, we consider a square lattice on a finite cylinder with a harmonic trap in the non-periodic direction, which should be a good approximation to an elliptical elongated trap.  The cylindrical system in Landau gauge has a ${\mathbb Z}_{L_x}$ symmetry along the circumference of the cylinder, when $\alpha_x=2\pi /L_x$.  With an appropriate construction of the Hamiltonian, a Fourier transform will turn the Hamiltonian into block diagonal form with each block corresponding to a certain momentum $k_x=p_xh/L_x=...,-h/L_x,0,h/L_x,...$ mod $h$ along the circumference of the cylinder.  These are the same momenta as in the Kac-Moody algebra with $k_x=k$. 

\begin{figure}[tbp]
  \begin{center}
      \subfigure{\includegraphics[width=84mm]{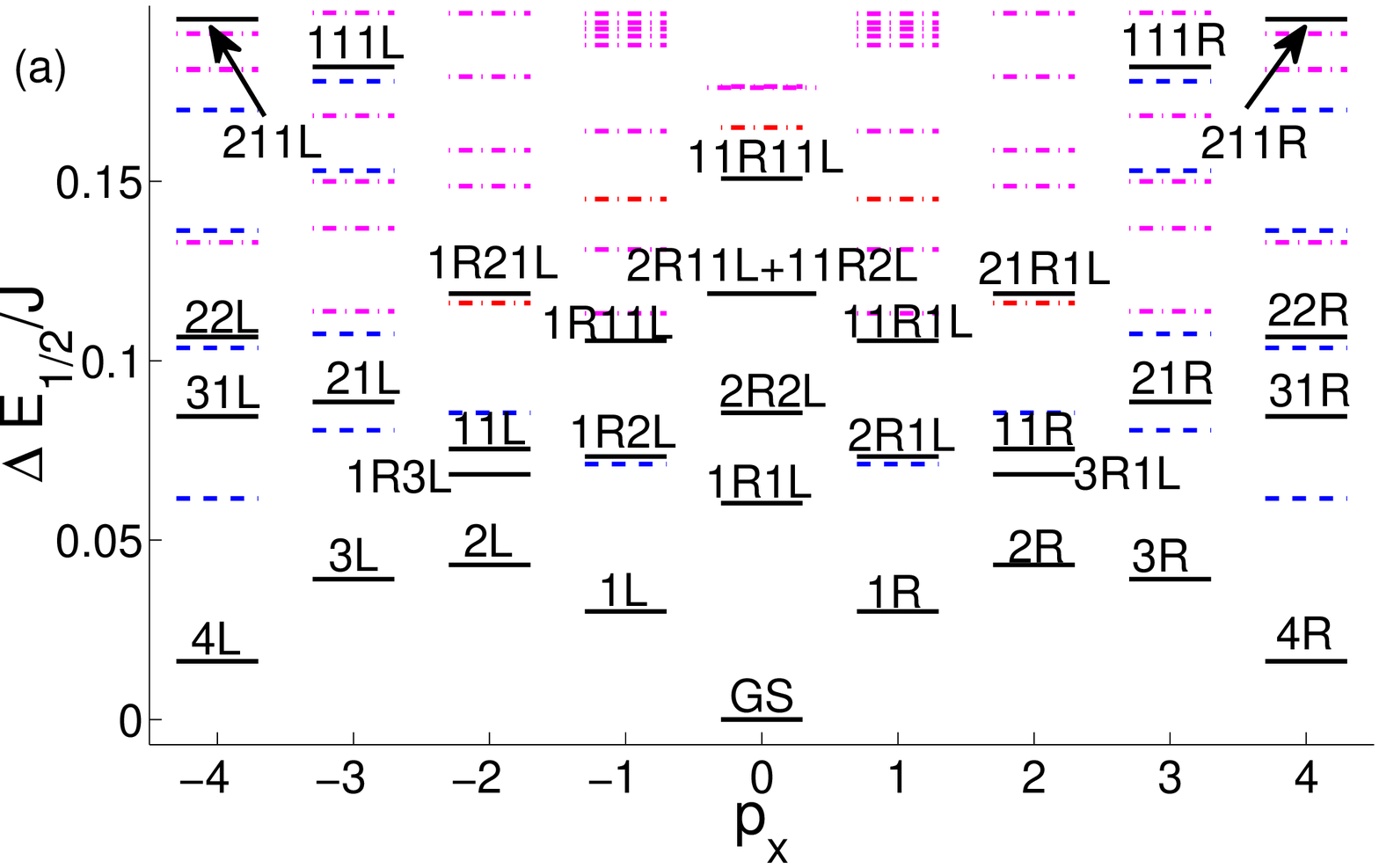}}\\
         \subfigure{\includegraphics[width=84mm]{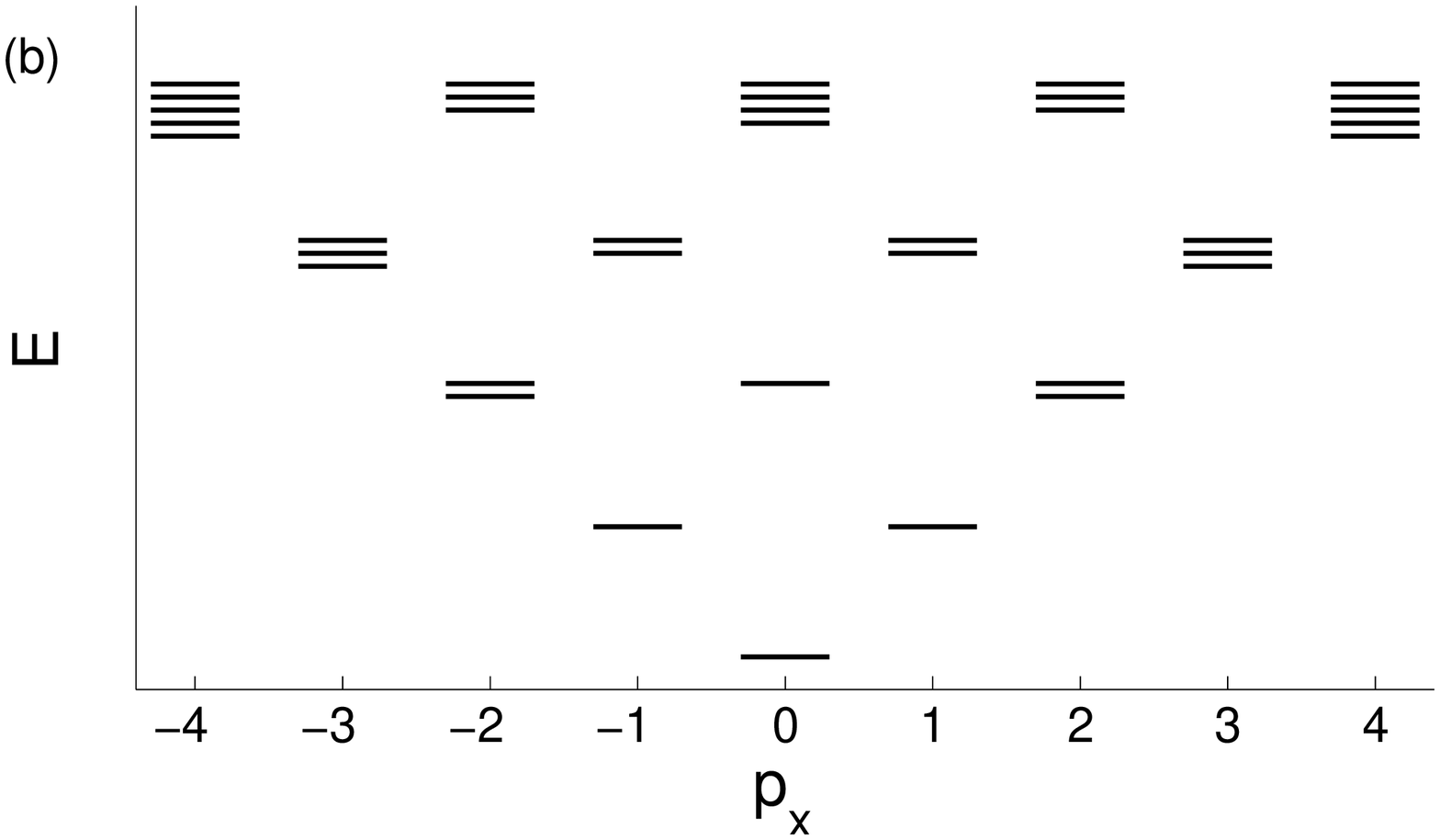}}
              \caption{(Color online) (a) Edge spectrum for $N=4$ particles at $\nu=1/2$ on a $9\times 15$ site cylinder in a harmonic trap as a function of momenta.  Apart from the expected edge spectrum (black solid lines), bulk excitations (red dash-dotted lines) and edge excitations from translated $\nu=1/2$ phases (blue dashed and magenta dashed-dotted lines, where the former are aliased to a lower momenta) are mixed in.  Two-fold degenerate states are marked with longer lines. (b) Analytic edge spectrum at $\nu=1/2$ for a large system.  Degenerate lines are drawn slightly apart for clarity.}
    \label{Cyl}
  \end{center}
\end{figure}

The edge spectrum on a cylinder in a 1-D harmonic trap $c_y/J=0.004$, is shown in Fig.~\ref{Cyl}(a). Three additional rows are required outside the ground state droplet on both edges to get all edge excitations with $|p_x|\le N$. No edge excitations with $|p_x|>N$ are found on any cylinder.  The edge states 111R1L, 1R111L, 1111R and 1111L are higher up in the spectrum and not shown. The discrepancy from the analytical spectrum, shown in Fig.~\ref{Cyl}(b), can be explained by two types of finite size effects. The lower-than-expected energy of the high momentum single particle modes depends on their overlap with the opposite edge, the increase is again from finite size effects especially in the furthest extending states.  Bulk excitations are shown with red lines. The bulk gap $\Delta E^B_{1/2}/J\approx 0.12$ is again larger than anticipated, $\Delta E^B_{1/2}/J\approx 0.08$ in Fig.~\ref{Intro}(b). 
      
We briefly mention a few results from other trap potentials.  The low-energy spectra on cylinders does not change qualitatively in other potential shapes.  
However, the spectra change completely on squares in circular traps $V_{\text{trap}}=c_r r^d$: the system shows a FQH edge for $1.5\lesssim d \lesssim 2.5$ with the best spectra at $d=2$, with $\Delta E_{\nu}\propto c_r/J$, for reasonable trap strengths. Increasing the trapping potential favors denser phases and the transitions to phases with higher filling fractions $\nu$ will hence occur at larger fluxes $n_{\phi}$. 

Lastly, we discuss which other FQH phases in optical lattices that potentially can be detected with our method.  The other phases in the bosonic Laughlin sequence $\nu=1/4,1/6,...$ cannot be the ground state without longer range repulsions in a trap~\cite{Hafezi}. 
In a trap, their ground state energies are higher than the energies presented here.  With decreasing magnetic field, we do not observe the higher order hierarchical states at $\nu=3/4,4/5,...$. But, we believe this is an artifact of the small systems studied.  The non-abelian phases in the Read-Rezayi sequence $\nu=1,3/2,2,5/2,...$~\cite{Read} are believed to be the ground state for bosons in a trap under certain conditions~\cite{Cooper-01,Rezayi}.  The $\nu=3/2$ phase is the simplest bosonic FQH phase that could be used for quantum computing~\cite{Freedman}.  In the other limit, at large flux per plaquette $n_{\phi}\approx 1/2$ $(1/3)$, are the lattice specific FQH phases believed to appear~\cite{Palmer,Moller,Hormozi}.  We considered lattices with uniform flux; recent work has shown convincingly that non-uniform magnetic fields can create the same FQH phases without Landau levels~\cite{Tang,Neupert,Sun,Sheng,Wang,Regnault}. The lessons in this work for trapping potentials and  geometries apply also to these more complex situations.  For completeness we mention a few recent proposals for detecting quantum Hall physics in ultracold gases other than by edge properties~\cite{Douglas,Zhao,Alba,Varney}

The main result of this work is that hard-core repulsion in a small simple lattice system of approximately square or circular geometry is sufficient to generate clearly resolved edge excitations for the bosonic FQH states at filling $\nu=1/2$ and $\nu=2/3$, provided that the conditions described above on flux per site and harmonic trap strength can be achieved; engineering flat or nearly flat bands is unnecessary.  In the cylindrical case, the edges are strongly interacting with each other and the edge excitations are much harder to distinguish, which suggests counterintuitively that increasing system size by going to an elongated trap may not be necessary or even desirable.  Observation of the bosonic FQH states discussed here would be a logical first step toward even more exciting new states that can be studied by similar methods.  The authors acknowledge helpful conversations with J. Guzman, G.-B. Jo, R.~S.~K.~Mong and S.~Parameswaran and support from the ARO Optical Lattice Emulator program.

\end{document}